# Efficient Sampling for Machine Learning Electron Density and Its Response in Real Space

Chaoqiang Feng[1,4], Yaolong, Zhang[2], Bin Jiang[3,4*]

[1]Hefei National Research Center for Physical Sciences at the Microscale, Department of Chemical Physics, University of Science and Technology of China, Hefei, Anhui 230026, China

[2]Department of Chemistry and Chemical Biology, University of New Mexico, Albuquerque, New Mexico 87131, United States

[3]Key Laboratory of Precision and Intelligent Chemistry, Department of Chemical Physics, University of Science and Technology of China, Hefei, Anhui 230026, China

[4]Hefei National Laboratory, University of Science and Technology of China, Hefei, 230088, China.

[*]Corresponding author: bjiangch@ustc.edu.cn

**Abstract**

Electron density is a fundamental quantity, which can in principle determine all ground state electronic properties of a given system. Although machine learning (ML) models for electron density based on either an atom-centered basis or a real-space grid have been proposed, the demand for the number of high-order basis functions or grid points is enormous. In this work, we propose an efficient grid-point sampling strategy that combines a targeted sampling favoring large density and a screening of grid points associated with linearly independent atomic features. This new sampling strategy is integrated with a field-induced recursively embedded atom neural network model to develop a real-space grid-based ML model for electron density and its response to an electric field. This approach is applied to a QM9 molecular dataset, a $H_2O$/Pt(111) interfacial system, and an Au(100) electrode under an electric field. The number of training points is found much smaller than previous models, when yielding comparably accurate predictions for the electron density of the entire grid. The resultant machine learned electron density model enables us to properly partition partial charge onto each atom and analyze the charge variation upon proton transfer in the $H_2O$/Pt(111) system. The machined learned electronic response model allows us to predict charge transfer and the electrostatic potential change induced by an electric field in an Au(100) electrode.

# 1. INTRODUCTION

Electron density, **n(r)**, is one of the most important and fundamental variables of physical and chemical systems. According to density functional theory (DFT), electron density can in principle uniquely determine all ground state properties of a given system, such as total energy, dipole moment, and higher-order multipoles[1]. Computing **n(r)** for a given configuration often requires to solve an electronic structure problem, *e.g.* by DFT. It is typically implemented in a self-consistent way and a reasonable estimate of an initial **n(r)** can in turn accelerate the solution of the Kohn-Sham (KS) equation in DFT[2]. Additionally, an accurate **n(r)** is a prerequisite for applying certain charge partitioning scheme to analyze the electron transfer in redox reactions or proton transfer processes.

Machine learning (ML) has become increasingly important in modern chemistry [3-9]. For instance, machine learned potentials (MLPs) have been widely used in molecular dynamics simulations of molecules, condensed phase materials, and interfacial systems[10-31]. Most of these MLPs learn the relationship between chemical structure and potential energy by generating atomic descriptors satisfies the translational, rotational, and permutational invariance with respect to energy and expressing total energy as a sum of atomic contributions for better scaling and generalizability[32, 33].

Electron density is inherently more information-rich than total energy so that a range of important electronic properties beyond total energy can be obtained directly from ML models that learn from electron density. Several machine-learned electron density (MLED) models have been developed[34-61]. In particular, **n(r)** is a three-dimensional probability distribution function in space whose value is dependent on the relative spatial position and the nuclear configuration. A key inductive bias for these MLED models is the choice of the representation of **n(r)**[62]. For example, Brockherde *et al.* and Bogojeski *et al.* first used a plane-wave basis representation of **n(r)** and independent kernel ridge models to regress individual basis function coefficients[34, 43]. This scheme has later been extended to predict excited-state electron density[63]. While the choice of a plane wave basis representation in these studies allows a systematic convergence of the electron density with increasing the number of basis functions, it is less efficient for anisotropic systems and not easily transferable to other molecules. Alternatively, Grisafi *et al.* chose to expand **n(r)** in a localized atom-centered basis functions[42, 44, 46] and developed a transferable model based on symmetry-adapted Gaussian process regression (SA-GPR)[36, 38, 49, 56, 59, 64]. Specifically, the atom-centered basis function is a combination of radial functions and spherical harmonics, whose corresponding coefficients are correlated with its local environment and learned by SA-GPR that preserves the correct symmetry of **n(r)**. This SA-GPR representation for electron density has been employed successfully in constructing MLED models for small molecules[64], non-covalent systems[36], condensed phases[49]. However, it requires an initial decomposition of the charge density onto predefined basis sets, which may introduce additional errors. Beatriz et al. have partially overcome this limitation by learning an optimal basis set together with coefficients to minimize the differences between the ML-predicted densities, which are obtained by projecting atomic bases onto a real-space grid, with the actual DFT charge densities[57]. In addition, Rackers *et al.* proposed an equivariant graph convolutional neural network (GCNN) to predict electron density of water clusters[53] and biomolecules[52]. It was found necessary to increase the order of the atom-centered basis set for accurate prediction, which led to a rapid increase of computational cost.

On the other hand, most DFT codes directly output electron density values in a real-space grid[65, 66]. As a result, one can conveniently learn these discrete values in the same spirit of learning potential energies and obtain a smooth function for $\mathbf{n}(\mathbf{r})$. The grid-based MLED models were first proposed by us in representing the embedded electron density to compute electronic friction at metal surfaces by viewing each grid point as a virtual atom[67, 68]. More recently, Jørgensen *et al.* constructed a grid-based equivariant message passing neural network (MPNN) model based on a uniform grid-point sampling scheme, referred to as DeepDFT, which exhibits superior accuracy than typical basis-based models[51]. Sunshine et al. further improved the DeepDFT model by enforcing charge balance and directly assigning zero density to grid points far away from any atoms[58]. The modified DeepDFT model was trained on a subset of the very large Open Catalyst 2020 (OC20) dataset[69] containing 56 different chemical elements, demonstrating the generalizability of grid-based MLED models.

In addition to learning electron density of isolated systems, describing the response of electron density to an external electric field, *i.e.* the charge difference with and without the field, is also of great importance, for example in electrochemistry. Very recently, Grisafi et al.[59] and Lewis et al.[70] independently developed similar ML models for electron density response (MLEDR) to a finite field based on the SA-GPR atomic basis representation for electron density[49]. Both models introduce the field-dependence to kernels and Grisafi et al. further include a long-distance equivariant descriptor to capture the non-local feature of the electron density response in metal electrodes[59]. Thus far, however, no real-space MLEDR models has been reported.

Indeed, a disadvantage of real-space grid-based MLED models is that they typically rely on a huge amount of data points, causing a substantial memory demand. This issue becomes more severe when the configuration space or the chemical space of the target system(s) is more complicated. To overcome this difficulty, Focassio et al. proposed a targeted sampling (TS) strategy which assigns a probability of point selection based on a normal distribution of the inverse of the charge density, which effectively samples more points with larger densities[60]. Based on this TS strategy, they were able to train an accurate model with just about 0.1% of the available grid points for individual molecular and material systems.

The purpose of this work is two-fold. The first goal is to further reduce the required number of grid points for training a grid-based MLED model to make it suitable for a large group of molecules and/or a complicated system involving a broad chemical and/or configuration space. The second goal is to extend the previously developed field-induced recursive embedded atom neural network (FIREANN) potential model[71-74] to describe electron density in the absence and presence of an electric field and thus its response. To this end, we leverage the concept of feature or structure selection in constructing MLPs[56, 75, 76] and propose an efficient two-step grid-point sampling strategy. First, the entire charge density mesh is sparsified using the density value-based TS. Second, the remaining points are screened based on their structural similarity in terms of associated grid-centered structural features. This sampling strategy is coupled with our FIREANN framework to construct an accurate MLED model using an extremely low fraction of data, 0.005%~0.015% of the entire charge density mesh. Numerical tests in the well-known QM9 molecular dataset and a liquid-solid interfacial $H_2O$/Pt(111) system validate this strategy and the resultant MLED model is found useful to serve as the foundation of the charge analysis. Moreover, the FIREANN model naturally introduces the field-dependence of charge density and the non-local effect, enabling an accurate description of the charge density response of an Au(100) electrode.

## 2. METHOD

## 2.1. Field-Induced Recursively Embedded Atom Neural Network Model for Electron Density

As mentioned above, **n(r)** is a three-dimensional spatial function, which in real-space is typically represented by discrete and scalar values on a dense grid in the simulation box in DFT codes. As proposed in our previous studies, the three-dimensional function **n(r)** of a system with $N$ atoms is formally analogous to a potential energy surface (PES) of a system with $N+N_g$ atoms where the extra $N_g$ atoms is ghost atoms[67, 68]. These ghost atoms have no physical meaning but represents the grid point position. As a result, the grid-centered local environments for both realistic atoms and the virtual atom can be described by conventional many-body atomic features, which are frequently used for constructing atomistic MLPs. Note that the local environment of ghost atoms is only related to the surrounding realistic atoms, and is mapped to the charge density or response values through atomic neural network. Here, we adopt the FIREANN approach[72, 73], which employs field-induced embedded atom density (FI-EAD) features to describe the atomic environment and system-field interactions[71]. Specifically, the $n$th FI-EAD feature of a central atom (including the virtual atom) $i$ is defined by the square of the linear combination of by all neighboring atomic orbitals and field-dependent orbital,

$$\rho_i^n = \sum_{l=0}^{L} \sum_{l_x,l_y,l_z}^{l_x+l_y+l_z=l} \frac{l!}{l_x!l_y!l_z!} \left[ \sum_{m=1}^{N_\varphi} d_m^n \left( \sum_{j \neq i}^{N_c} c_j \varphi_{l_x l_y l_z}^m(\vec{r}_{ij}) + c_\varepsilon \varphi_{l_x l_y l_z}(\vec{\varepsilon}_i) \right) \right]^2. \quad (1)$$

where $N_c$ is the number of neighbor atoms, $c_j$ is the combination coefficient of the $j$th neighbor atom, and its corresponding atomic orbital is obtained by the contraction (the second linear combination in Eq. (1)) of $N_\varphi$ primitive Gaussian-type orbitals (GTO, $\varphi_{l_x l_y l_z}^m(\hat{r}_{ij})$) with $d_m^n$ being the contraction coefficient of the $n$th FI-EAD feature and $m$th GTO. A primitive GTO at atom $j$ is characterized by its center ($r_s$), width ($\sigma$), and angular moment ($l=l_x+l_y+l_z$) as,

$$\varphi_{l_x l_y l_z}^m(\hat{r}_{ij}) = (x_{ij})^{l_x} (y_{ij})^{l_y} (z_{ij})^{l_z} \exp\left[-\frac{(r_{ij}-r_s)^2}{2\sigma^2}\right] f_c(r_{ij}). \quad (2)$$

where $\hat{r}_{ij} = \hat{r}_i - \hat{r}_j$ is the relative position vector, with $r_{ij}$, $x_{ij}$, $y_{ij}$, and $z_{ij}$ being its norm and three Cartesian components, $f_c(r_{ij})$ is a cosine-type cutoff function. The virtual field vector-dependent is defined as,

$$\varphi_{l_x l_y l_z}(\vec{\varepsilon}) = (\varepsilon_x)^{l_x} (\varepsilon_y)^{l_y} (\varepsilon_z)^{l_z}. \quad (3)$$

Note that in practice the summation order is exchanged in Eq. (1) for faster evaluation.

Varying these hyperparameters forms an array of FI-EAD features, which encode three-body interactions implicitly by the summation of $l > 0$ terms with the $\sim O(N_c)$ scaling. To incorporate higher-order and more non-local interactions, $c_j$ can be also the output of an atomic NN that depends on the $j$th atom's local environment and can be updated iteratively, namely,

$$c_j^t = g_j^{t-1}\left[\boldsymbol{\rho}_j^{t-1}\left(\mathbf{c}_j^{t-1},\boldsymbol{r}_j^{t-1}\right)\right] \quad (4)$$

where $g_j^{t-1}$ represents the $j$th atomic NN module in the $t$th iteration. This leads to a message-passing NN architecture, which has been successfully applied to learn PESs and response properties of molecular and condense phased systems[72, 73]. It should be noted that the learning target of MLED model is the electron densities in absence of an electric field, and that of MLEDR model is defined as the difference between the perturbed electron densities and the electron densities of the corresponding isolated electrodes, i.e., $\Delta n_e = n_e - n_0$.

Note that the FI-EAD feature vector here is used to describe the atomic environment and does not represent the true electron density of the atoms system. For clarity, we use $\rho(\{\rho\})$ to denote the FI-EAD features (or set), and $\boldsymbol{n}$ to denote the charge density of systems. The corresponding hyper-parameters of FI-EAD features are optimized iteratively during the cycle of sampling and training.

## 2.2. Linearly Independent Feature-based Grid-point Sampling

A remarkable difference between training a PES and $\mathbf{n}(\mathbf{r})$ in real space is that every single configuration on the PES provides a dense three-dimensional grid of density data.

In practice, this leads to highly redundant data points and using the entire mesh is neither efficient nor necessary. Because $\mathbf{n}(\mathbf{r})$ is near zero in a wide range of space, *e.g.* a position far from any nuclei, a random sampling (RS) of grid points would sample too many near-zero density values, leading to inefficient coverage of the most probable region of $\mathbf{n}(\mathbf{r})$. In this regard, Jørgensen *et al.* randomly sampled 1000 grid points for each configuration to train their equivariant MLED model and found the removal of some low-density points can be beneficial[51]. Later, Focassio *et al.* proposed a probabilistic TS procedure that tends to sample more grid points with high densities, by which an accurate MLED model can be constructed with just 0.1% of the entire mesh[60]. However, significant redundancy remains in the resulting dataset as many of these grid points would have similar local environments due to symmetry, especially for small molecules with high symmetry and crystals. As a result, we propose to select only these grid points which have their atomic features linearly uncorrelated with others. A similar strategy was previously applied to select a minimal number of linearly independent atomic features that best represent the local environment in constructing EANN potentials[76].

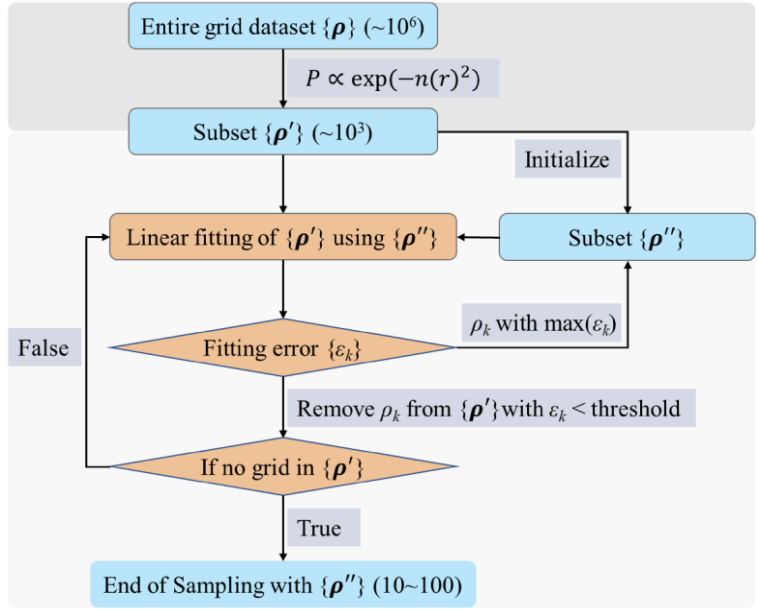

Figure 1: A schematic workflow of the proposed two-step grid-point sampling strategy.

The proposed two-step grid-point sampling workflow is illustrated in Fig. 1. The three-dimensional grid generated by DFT typically consists of a few million points, which is our initial dataset. We first perform the TS strategy of Focassio *et al.* to assign a higher probability for sampling points with large densities[60], For a grid point candidate $r_g$, the relative sampling probability is given by,

$$P(r_g) = \frac{1}{\sqrt{2\pi}} \exp\left(-\frac{(1/\mathbf{n}(r_g))^2}{2\sigma^2}\right), \qquad (5)$$

where the hyperparameter $\sigma$ controls the width of the probability distribution. The smaller $\sigma$, the higher concentration of sampled grid points is in the high-density region. After collecting a subset with say $N_g$ grid points by TS, we calculate their grid-centered FI-EAD features and perform a second-step linearly independent feature-based sampling (LIFS) to judge their similarity. The first point is randomly selected out from the subset {*ρ'*} and add into the {*ρ"*}, then every later candidate's FI-EAD feature vector will be linearly fitted by these feature vector(s) of the selected data point(s) and only candidates with linearly independent features are accepted and added into {$\boldsymbol{\rho}''$}. For example, for the $k$th candidate, the deviation of the linear least-square by feature vectors of these accepted points ($\varepsilon_k$) is normalized to be used as the criterion of linear independence of the candidate's feature vector $\boldsymbol{\rho}_k$,

$$\varepsilon_k = \frac{\left\|\boldsymbol{\rho}_k - \sum_{i}^{N_s} \boldsymbol{\rho}_i \gamma_i \right\|}{\left\|\boldsymbol{\rho}_k\right\|}, \qquad (6)$$

where $\gamma_i$ represents the least-square coefficient. A candidate with a large $\varepsilon_k$ implies that its atomic feature is dissimilar to these of selected points and should be thus added. In contrast, the grid point with a small $\varepsilon_k$ indicates that its atomic feature can be well represented by the linear combination of other features and thus not necessary. This procedure can be performed iteratively until all points with linearly independent features are sorted out. In practice, we find that on average merely 10~100 grid points for each nuclear configuration are necessary, which made it possible to train an MLED model for a complex system or a mix of many subsystems with an affordable cost.

## 2.3. Datasets and training setup

We have tested the performance of the proposed sampling strategy in learning electron density with two distinct datasets. The first one is the QM9 dataset[77, 78], which contains 133,885 small organic molecules and is widely used for benchmarking ML models for predicting molecular properties. The electron density of this QM9 dataset were taken from previous work[51, 79]. The second one is a liquid-solid interfacial $H_2O$/Pt(111) system, where the surface was represented by a four-layer Pt(111) periodic slab using a (4×4) supercell, initially separated by a 18Å vacuum space in the vertical direction. The electrochemical interface was then built up by fully filling the vacuum space between repeated slabs with 60 water molecules. A total of 1000 configurations were generated by running ab initio molecular dynamics (AIMD) simulations with a thermostat temperature of 300 K. For both datasets the electron densities were calculated with Vienna Ab initio Simulation Package (VASP)[80, 81], and the Perdew-Burke-Ernzerhof (PBE) density functional[82] and projector augmented wave (PAW) method[83] were employed. The wave function of valence electrons was expanded using plane waves with energy cutoff of 400eV and 600eV for the QM9 dataset and the $H_2O$/Pt(111) dataset, respectively. The first Brillouin zone was sampled only at the Gamma center for QM9 dataset, and with a 3×3×1 Monkhorst-Pack k-point mesh for the $H_2O$/Pt(111) dataset[84]. The charge densities were output on a three-dimensional grid evenly spaced by 0.08~0.10 Å. Besides, the charge response dataset consists of 260 slab configurations of Au(100) were taken from previous work[59], with 3~17 metal layers in a (2×2) supercell. The charge responses were computed with the open source CP2K package with the PBE density functional, Goedecker-Teter-Hutter (GTH) pseudo-potentials, and double-ζ basis sets with one set of polarization functions (DZVP)[66, 85-89]. The charge transfer at the two metal surfaces was calculated under a uniform electric field along z, and the field intensity was set to -1.0 V/Å.

Note that the average number of grid points per configuration is ~0.7 million for the QM9 dataset and ~5.4 million for the $H_2O$/Pt(111) dataset. In practice, before sampling grid points for training MLED models, 1000 molecules from the QM9 dataset and 100 configuration from the $H_2O$/Pt(111) dataset were selected randomly as the test set, respectively. We performed the TS in the residual configurations to select 1000 grid points per molecule in the QM9 dataset and 5000 grid points per configuration in the $H_2O$/Pt(111) dataset with a width parameter $\sigma$ of 30 and 80 $Å^3e^{-1}$, respectively. A subsequent LIFS was performed from these collections of grid points. The final dataset sampled by this two-step procedure were divided into the training and validation sets with a ratio of 9:1. The training process was optimized using the AdamW algorithm[90]. The learning rate, initially set to 0.001, decayed by a factor of 0.5 whenever the validation error did not decrease for 50 epochs. Training was stopped when the learning rate dropped below $1\times10^{-6}$. More information about the architecture of the FIREANN model is given in Table 1.

**Table 1:** NN structures (the number of neurons in each hidden layer), cutoff radii (in Å), maximum angular momentum, the number of message passing iterations, batch size and the number of FI-EAD features used in training FIREANN electron density models for QM9 and $H_2O$/Pt(111) datasets and electron response model for Au electrode dataset.

| Parameter | System | | |
|---|---|---|---|
| | QM9 | $H_2O$/Pt(111) | Au electrode |
| NN structure for atomic properties | 32×32 | 64×64 | 128×128 |
| NN structure for orbital coefficients | 16×16 | 32×32 | 64×64 |
| Cutoff radius/Å | 4.0 | 4.0 | 6.0 |
| Maximum angular momentum | 2 | 2 | 2 |

| Parameter | System | | |
|---|---|---|---|
| | QM9 | $H_2O$/Pt(111) | Au electrode |
| Number of messages-passing iterations | 2 | 2 | 6 |
| Batch size | 16 | 8 | 8 |
| Number of features | 108 | 108 | 234 |

## 3. RESULTS and DISCUSSION

### 3.1. Performance of the MLED Model

### 3.1.1 QM9 dataset

Let us first discuss the prediction accuracy of the FIREANN models for $n(r)$ of molecules in the QM9 dataset that cover a vast chemical space. The proposed TS+LIFS strategy was used to sample ~100 grid points per molecule, collecting about 13 million grid points. Compared to the entire grid of representing $n(r)$ for all molecules in the dataset, the FIREANN model was thus trained with only about 0.014% of available grid points. The correlations between DFT calculated electron densities and FIREANN predictions for different datasets are illustrated in Figure 2. Note in the test set that grid points are more distributed in the low-density region, which is a natural consequence of the even distribution of grid points and a large area filling with near-zero electron density. Indeed, more than half of grid points in the QM9 dataset have charge densities lower than $10^{-3}$ e·Å$^{-3}$. While the TS algorithm favors more high-density points, the LIFS algorithm excludes these grid points with similar local environments, and as a result, the charge densities in the training set and validation set become more evenly distributed in the whole range between 0 and 14 e·Å$^{-3}$. The optimized FIREANN model yields a very low root-mean-square error (RMSE) of 0.0011 e·Å$^{-3}$ for the test set with an extremely high correlation coefficient ($R^2$=0.999986). This is even lower than the RMSE for the validation set (0.0016 e·Å$^{-3}$) and for the training set (0.0019 e·Å$^{-3}$), indicating that the FIREANN model can accurately predict **n(r)** in the entire mesh, including low-density regions that are uncommon in the training dataset.

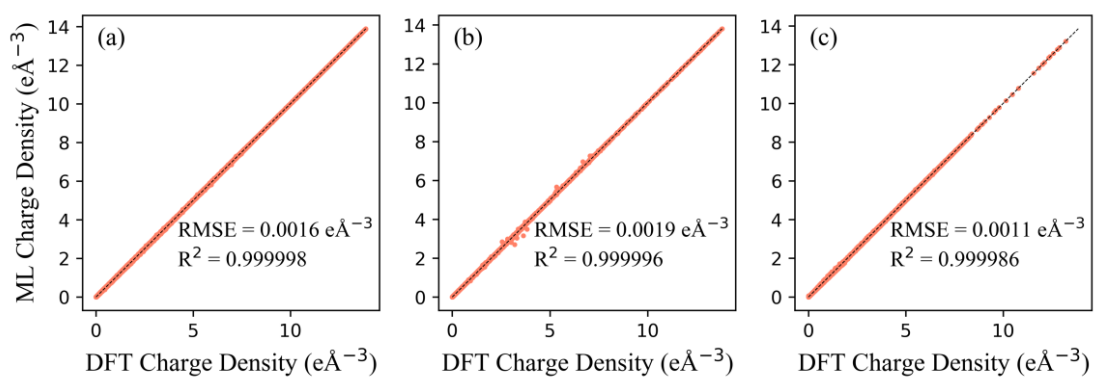

**Figure 2.** Correlation between DFT and ML charge densities in the training set (a), validation set (b), and test set (c) as part of the QM9 dataset. The black dashed lines represent perfect linear correlation.

Figure 3 displays the error iso-surfaces for several representative molecules in the QM9 dataset. These molecules comprise common chemical elements (i.e., Carbon, Hydrogen, Oxygen, and Nitrogen) and prevalent chemical interactions in the QM9 dataset. The prediction error for each molecule in the entire simulation box is evaluated with the normalized mean absolute error (MAE), which was introduced by Jørgensen and Bhowmik[51] and defined as,

$$\varepsilon_{mae} = \frac{\int_{r \in V} |n_{\text{DFT}}(r) - n_{\text{NN}}(r)|}{\int_{r \in V} |n_{\text{DFT}}(r)|}, \tag{7}$$

where $n_{\text{DFT}}(r)$ represents the DFT electron density and $n_{\text{NN}}(r)$ is the FIREANN predicted electron density at the grid. As seen in Figure 3, large prediction errors primarily appear around atoms or chemical bonds, *i.e.* in the regions with large charge densities. This indicates that sampling less data points in low-density regions does not reduce the prediction ability of the FIREANN model there. Additionally, larger molecules have generally smaller normalized MAEs, as they possess a higher average charge density.

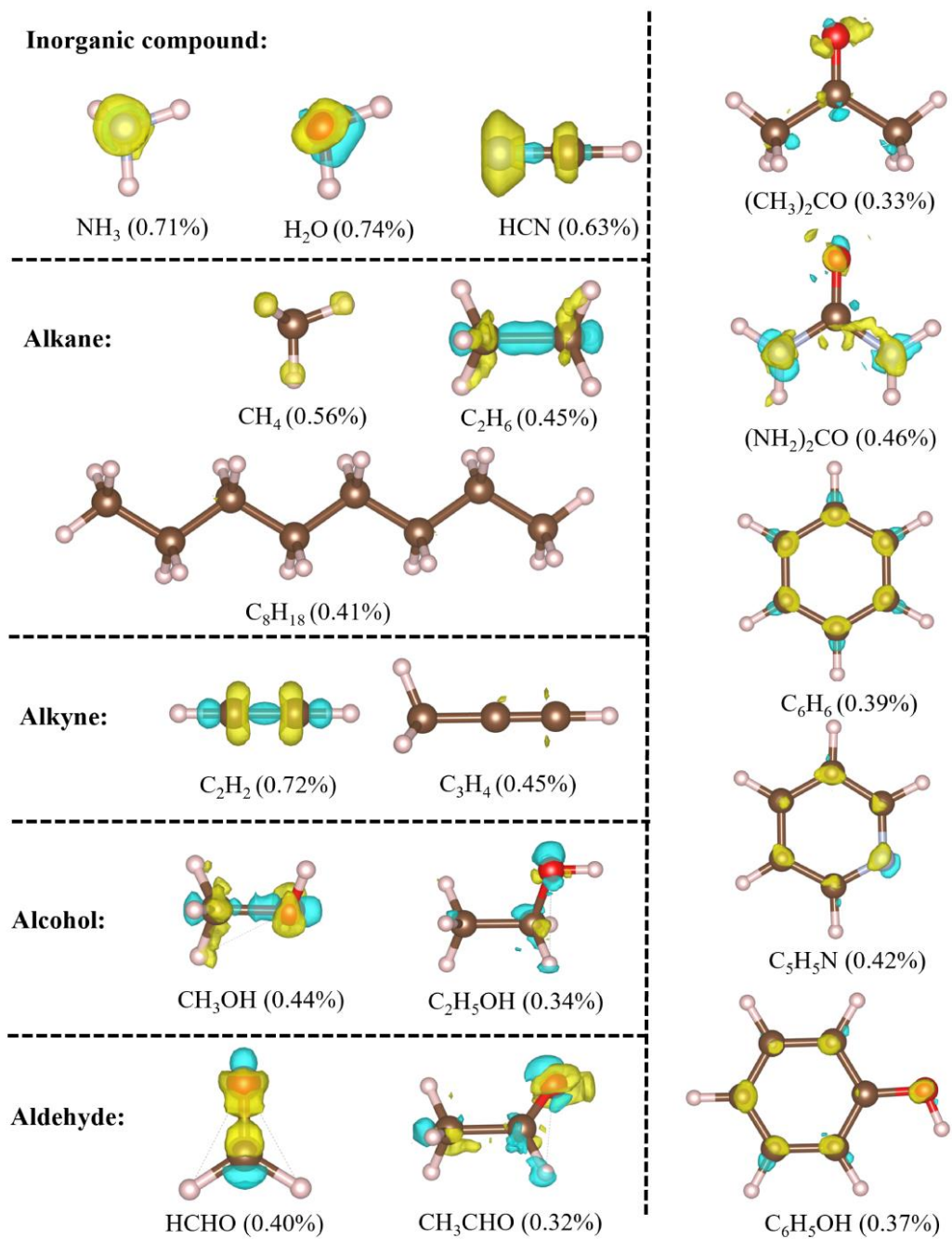

**Figure 3.** Prediction error iso-surfaces ($\pm 0.001$ e·Å$^{-3}$) for several exemplary molecules in QM9 dataset containing H (white), C (brown), O (red), and N (grey) atoms. Numbers in parentheses denote the normalized MAE for each molecule, as defined in Eq. 6.

In Figure 4, the accuracy of the FIREANN model is further investigated through a scan along a central line across the propine and phenol molecules. The variation of charge density at the grid along this line is essentially indistinguishable by DFT and FIREANN. The maximum difference between them never exceeds 0.02 e·Å$^{-3}$, which is ~1% of the maximum density. It should be noted that the ML model here only learns the density of valance electrons as core electrons are described by pseudopotentials. In some circumstances, for example, the aggressive pseudopotential removes some charges from carbon atomic centers, corresponding to some local minima in the charge density variation in Figure 4. The total electron density can be in principle recovered by adding the core charge to the ML-predicted valence charge density.

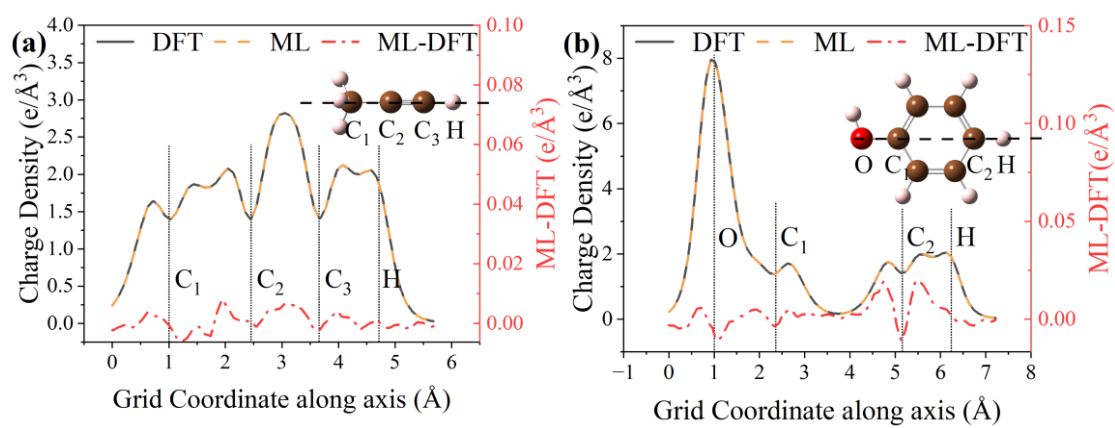

**Figure 4.** DFT and ML-predicted charge density for propine and phenol molecules computed along the line indicated in the inset, the prediction error is also provided on the right-hand side scale (red).

### 3.1.2 H$_2$O/Pt(111) dataset

Next, we turn our attention to the liquid-solid H$_2$O/Pt(111) interfacial system. Compared to the QM9 dataset, this system contains many more electrons and a periodic electron density distribution. Additionally, the H$_2$O/Pt(111) dataset involves more complex chemical interactions, including the metallic bonds within the platinum slab and hydrogen bonds between water molecules. Using a similar procedure as for the QM9 dataset, the TS+LIFS strategy selected only 200 grid points out of the raw ~500 million grid points for each configuration generated by VASP, yielding a dataset with 90,000 grid points and representing ~0.004% of the entire mesh. The RMSEs for the training set and validation set are 0.0031 and 0.0036 e$Å^{-3}$, respectively, corresponding to ~1% of the average density value. Similar to the results of QM9, the RMSE for the test set (0.0026 e$Å^{-3}$) is slightly smaller than that in both the training and validation sets, as low-density data have a relatively larger proportion in the former. Figure 5 shows a perfect agreement between the DFT and FIREANN predicted charge density distributions of a representative configuration of $H_2O$/Pt(111) in the test set, where that the maximum prediction error is less than 0.1 e·$Å^{-2}$, ~0.5% of the maximum integrated density along the X-axis.

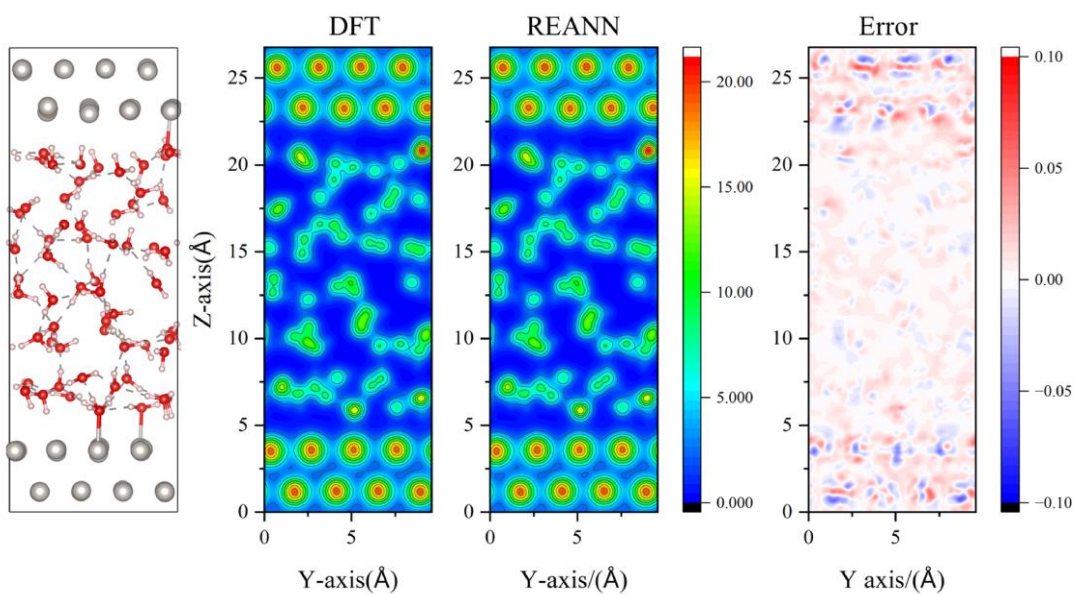

**Figure 5.** Comparison of electron density of an exemplary configuration in the test set of $H_2O$/Pt(111) obtained from DFT and FIREANN. From Left to right: (a) A snapshot of the $H_2O$/Pt(111) trajectory; Contour plots of DFT calculated (b) and FIREANN

predicted (c) electron density integrated along the X axis; (d) the difference between (c) and (d). The units of color bar are e·Å$^{-2}$.

## 3.2 Predicting the Electron Density Response

To validate the capability of the TS+LIFS strategy and the FIREANN model in predicting the electron density response, we trained the MLEDR model for an Au(100) electrode. It is worth noting that the electron density response of a conducting system to an external field exhibits a remarkable nonlocal behavior, as exemplified by the Au electrode system, leading to an accumulation of opposite charges on the two sides of the metal electrode. This renders the use of six message-passing iterations in the training process, as listed in Table 1.

Figure 6 compares the trained FIREANN model for electron density response of the electrode with the corresponding DFT profile and the corresponding electrostatic potential (ESP) as a function of the vertical position of the Au(100) slab. The charge transfer between the two sides of the metal electrode is obtained by integrating the charge response from the central layer of the slab to the left or right vacuum regions. Our FIREANN model precisely reproduces the charge transfer of 0.1871 *e* between the two sides of electrode, in agreement with the DFT value of 0.1869 *e*. The variation of the ESP is computed directly from the predicted charge density response. We find that the slope of the averaged ESP is approximately -1.0 V/Å inside the metal electrode, corresponding exactly to the field strength. This indicates that the internal electric field generated by the response perfectly screens the opposing external field. This result is very similar to that predicted by the SA-GPR model plus a long-distance equivariant descriptor for the same system[59].

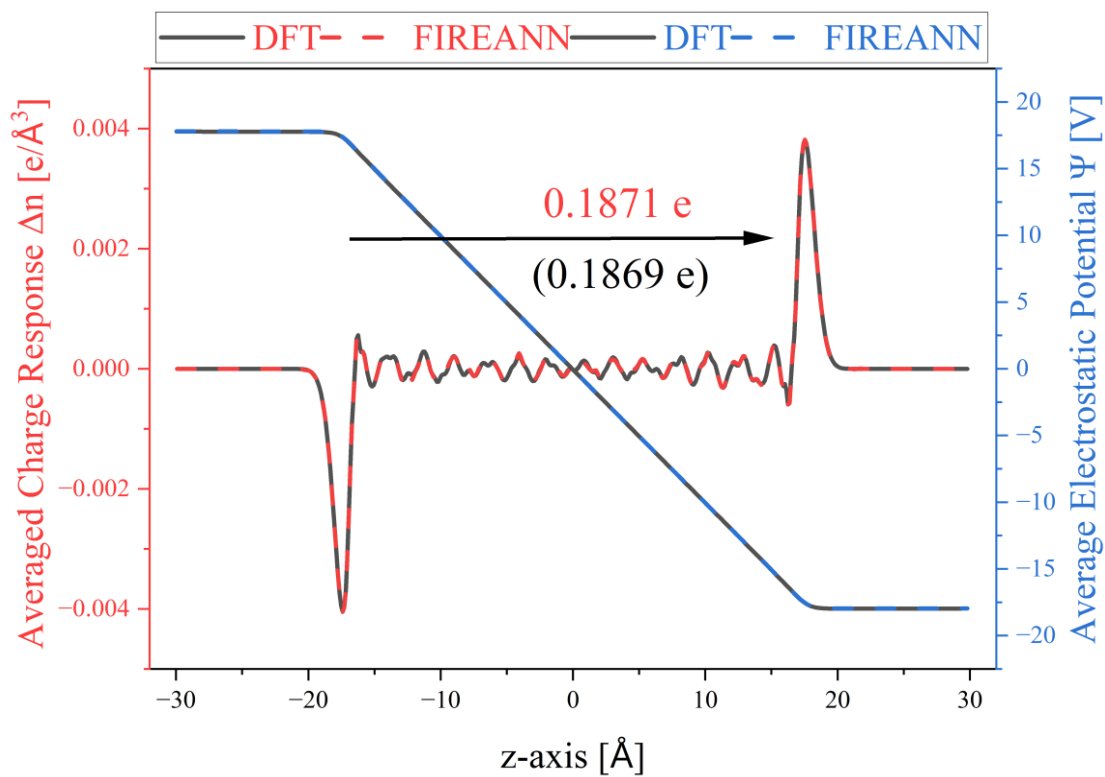

**Figure 6.** Response of charge density and electrostatic potential averaged over the *xy* plane of an Au (100) electrode under an external electric field of $E_z$ = -1 V/Å.

### 3.3 Comparison of sampling strategies

We further examine to what extent the LIFS strategy enhances the data efficiency. In Figure 7, we compare the learning curves (the normalized MAE *versus* the size of the training set) derived from the TS and TS+LIFS strategies. The learning curve for the RS strategy was not shown here, as previous work has already proven that the RS strategy exhibits much worse efficiency and accuracy[60]. For the QM9 dataset, the normalized MAE is again evaluated over the entire grid for outputting n(r) of randomly chosen 1,000 molecules. As shown in Figure 7 (a), the MAE decreases quickly with more grid points added in the first a few data selection cycles. Impressively, the same level of accuracy is achieved by the TS+LIFS strategy using less than 40 grid points from each molecule, compared with that of the TS strategy alone using 400 points. This result highlights the superior data efficiency of the former method. A similar numerical experiment for the $H_2O$/Pt(111) dataset shown in Figure 7(b) also suggests that the MAE of the TS+LIFS strategy is always lower than that of the TS strategy for a given number of grid points. The former quickly converges with 100 grid points, while the latter is still slowly decreasing with 200 points. In this case, the TS+LIFS strategy selecting 40 points can reach a comparable MAE to that by the TS strategy alone selecting 1000 points.

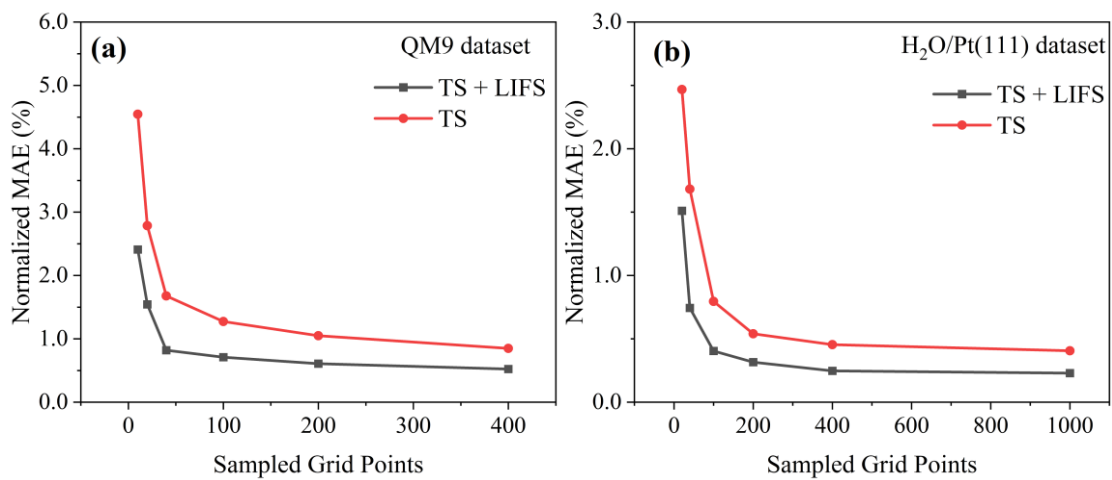

**Figure 7.** Learning curves of the FIREANN model for electron density in the QM9 (a) and $H_2O$/Pt(111) datasets (b) using different grid point sampling strategies, *i.e.* the normalized MAE as a function of the average number of grid points sampled for each molecule (or configuration)

To further understand the effectiveness of the LIFS strategy, we visualize the distribution of grid points sampled by different strategies in the feature space in Figure

8. Specifically, using the test set of QM9 as an example, 30,000 representative grid points were sampled using the RS, TS, and TS+LIFS strategies, respectively. The high-dimensional feature space is reduced by principal component analysis (PCA) and two most important components are chosen for the visualization. As shown in Figure 8(a), the RS strategy tends to sample more points located in the low-density region, as the probability of sampling each point in the real space is equal and the low-density region is more broadly distributed. Consequently, it is necessary to collect an overwhelming number of points when applying the RS strategy in order to cover the chemical space more evenly and ensure the model accuracy, typically $10^3$ to $10^4$ grid points per molecule. In comparison, the large-density region is partially covered by the TS strategy, as shown in the upper-right blue region of panel (b), since it offers a higher probability of sampling grid points with large electron densities. One step further, the TS+LIFS strategy allows us to sample grid points with linearly-independent features only, or in other words, remove those ones with the similar local environments, thus yielding a much more unbiased distribution of the sampled grid points in the feature space. Therefore, the dataset size is significantly reduced to train a balanced MLED model using this sampling scheme.

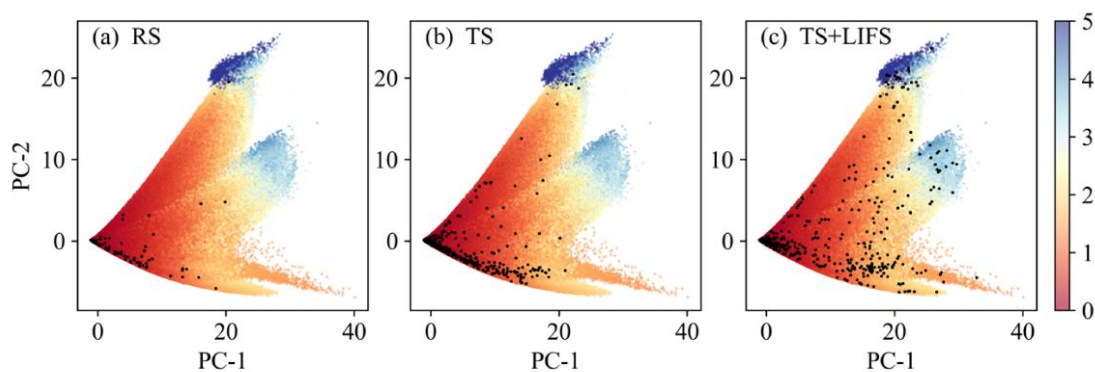

**Figure 8.** Comparison of the distribution of data points sampled by the RS (a), TS (b) and TS+LIFS (c) strategies in the reduced feature space for the QM9 test set (1000 randomly chosen molecules). Dimension reduction of the EAD features is performed by the principal component analysis (PCA) algorithm. For visual clarity, the electron density value is truncated between 0 and 5 eÅ$^{-3}$ and only a random subset (1%) of the grid points is depicted.

### 3.4 Atomic Charge Predictions

A well trained MLED model is supposed to be used to determine atomic charges, which are often useful to analyze electron transfer processes in chemical reactions. One common charge partitioning scheme is the so-called Bader charge analysis, which attributes partial charges to different atoms separated by zero-flux surfaces of electron density[91-94]. To validate our MLED model, Bader charge analysis has been performed with DFT-based and our FIREANN-based electron density distributions for a series of $H_2O$/Pt(111) configurations involving a proton transfer from one water molecule to another.

Figure 9 compares the Bader charges of the hydrogen atom moving from the donor to the acceptor water molecule. For simplicity, the distance between the oxygen atoms in the two relevant water molecules (O1-O2) is fixed at 2.50 Å, where the shorter $O_1$-H bond in the donor is initially ~0.96 Å and the longer hydrogen bond length (O2-H) is ~1.54 Å. Interestingly, the Bader charge of the transferring hydrogen atom is constantly zero as the O1-H bond moderately elongates. This is because the electron density of the donor molecule is mostly assigned to the oxygen atom given its high electronegativity, making the hydrogen atom virtually like a proton. However, as the O1-H distance exceeds ~1.15 Å, roughly the sum of the covalent radii of isolated oxygen and hydrogen atoms, a zero-flux surface of the electron density appears between O1 and H. Consequently, a partial charge starts to be assigned to the hydrogen atom (0.35 e), which slightly increases to a maximum (~0.39 e) when the hydrogen atom lies in the middle of the two oxygen atoms and then slowly decreases as the hydrogen atom further moves to the acceptor water molecule. Symmetrically, when the O2-H distance becomes shorter than ~1.15 Å, the zero-flux surface disappears suddenly so that the electron density of the acceptor is now assigned to O2, making the atomic charge of the transferred H atom zero again in the end of this proton transfer process. Since the zero-flux surface is sensitive to the actual electron density distribution, this result clearly demonstrates that the FIREANN model precisely captures the variation of **n(r)** upon the breaking and forming of hydrogen bond.

Note that the integration of the charge density over the space predicted by the ML model differs from the number of valence electrons by less than 0.1%, although the charge balance is not enforced in our FIREANN model. This slight charge imbalance has little impact on the Bader charge analysis results. However, in some physical models of periodic systems, it is problematic if the total charge is not exactly balanced[58]. In such cases, charge balance can be reinforced by scaling the predicted charge density.

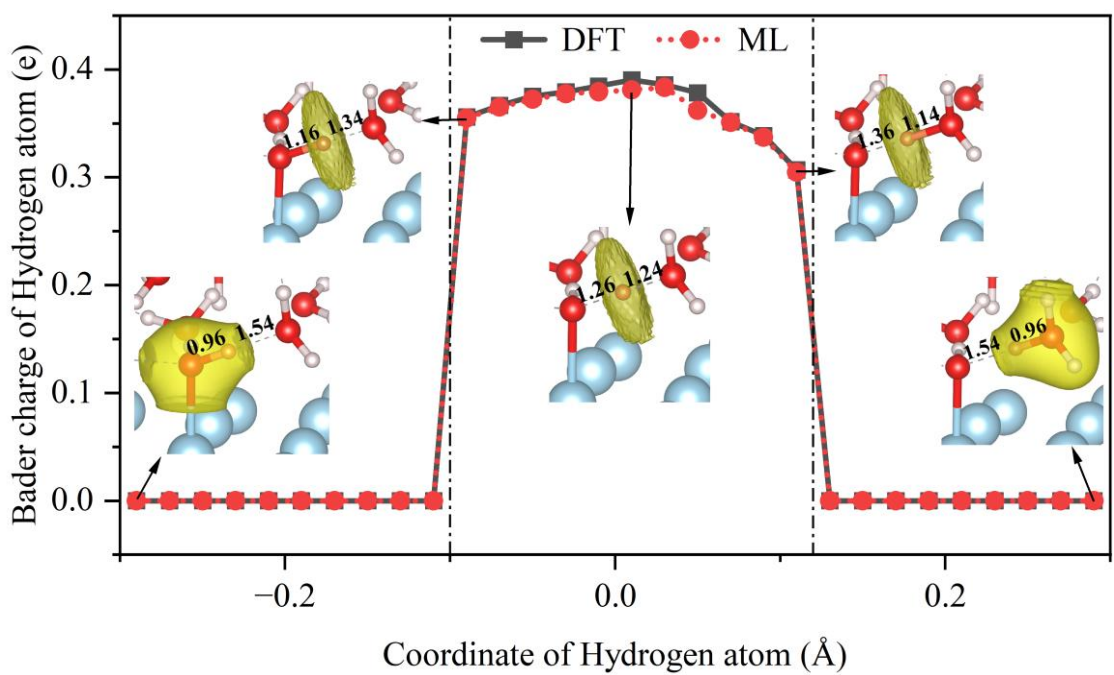

**Figure 9.** Variation of the Bader charge of the hydrogen atom during a proton transfer process in the H$_2$O/Pt(111) system. Bader charge analysis is performed using charge densities obtained from DFT (black) and the FIREANN model (red). Five representative structures in the proton transfer process are shown with H (white), O (red), Pt (blue) atoms. The two most relevant O-H distances in these configurations are marked in Å. The horizontal axis represents the displacement of the hydrogen atom relative to the midpoint between the two oxygen atoms.

## 4. CONCLUSIONS

In this work, we propose a very efficient hybrid sampling of grid points to train machine learning electron density and response models in the real space. This strategy starts from a value-based targeted sampling which favors grid points in the high-density region, followed by removing these points associated with linearly dependent atomic features which are regarded as having similar local chemical environments. Combining the proposed strategy and our field-induced recursively embedded atom neural network method allows us to obtain an accurate electron density or response representation using merely about 0.005%~0.015% of the entire grid of charge density generated by density functional theory codes, in both the QM9 molecular dataset and a periodic $H_2O$/Pt(111) interfacial system. Moreover, our model is also able to accurately predict the non-local charge transfer in an Au(100) electrode under an applied electric field. The resulting machine learning electron density model has been used to perform Bader charge analysis and investigate the electron transfer involved in the proton transfer process of $H_2O$/Pt(111). This study suggests that the linearly-independent feature selection is a general way to efficiently select not only atomic features but also data points to train machine learning models. Combining with machine learned potentials, the proposed machine learning electron density model may be applied, for example, to electronic friction based non-adiabatic molecular dynamics simulations on metal surfaces[95], to predict the charge density response in electrochemical cells and perform constant-potential simulations of electrochemical systems with DFT accuracy[59]. Further applications in these scenarios are underway in our group.

## Conflicts of interest

The authors declare no competing financial interest.

## Acknowledgement

This work is supported by the Strategic Priority Research Program of the Chinese

Academy of Sciences (XDB0450101), Innovation Program for Quantum Science and Technology (2021ZD0303301), the CAS Project for Young Scientists in Basic Research (YSBR-005), the National Natural Science Foundation of China (22325304, 22221003, and 22033007). We acknowledge the Supercomputing Center of USTC, Hefei Advanced Computing Center, Beijing PARATERA Tech CO., Ltd. for providing high-performance computing service.